\documentclass[twoside]{mhd}
\usepackage{graphicx}
\usepackage{amsmath}
\usepackage{amssymb}
\usepackage{graphicx}

\mhdhead{40}{1}{1}

\title{Nonmodal analysis of helical and azimuthal magnetorotational instabilities}

\author{G. Mamatsashvili\inst{1}\inst{,2}\inst{,3} and F. Stefani
\inst{1}}
\institute{Helmholtz-Zentrum Dresden-Rossendorf, P.O. Box 510119,
D-01314 Dresden, Germany \and Department of Physics, Faculty of
Exact and Natural Sciences, Tbilisi State University, Tbilisi 0179,
Georgia \and Abastumani Astrophysical Observatory, Ilia State
University, Tbilisi 0162, Georgia}

\begin{document}
\maketitle


\begin{abstract}
The helical and the azimuthal magnetorotational instabilities
operate in rotating magnetized flows with relatively steep negative
or extremely steep positive shear. The corresponding lower and upper
Liu limits of the shear, which determine the threshold of modal
growth of these instabilities, are continuously connected when some
axial electrical current is allowed to pass through the rotating
fluid. We investigate the nonmodal dynamics of these instabilities
arising from the nonnormality of shear flow in the local
approximation, generalizing the results of the modal approach. It is
demonstrated that moderate transient/nonmodal amplification of both
types of magnetorotational instability occurs within the Liu limits,
where the system is stable according to modal analysis. We show that
for the helical magnetorotational instability this
magnetohydrodynamic behavior is closely connected with the nonmodal
growth of the underlying purely hydrodynamic problem.
\end{abstract}


\section*{Introduction}

The helical and azimuthal magnetorotational instabilities are
dissipation-induced instabilities that have attracted growing
theoretical and experimental interest in recent years. They operate
in magnetized shear flows with high resistivity, or very low
magnetic Prandtl numbers, ${\rm Pm}=\nu/\eta\ll 1$, i.e., the ratio
of viscosity $\nu$ to magnetic diffusivity $\eta=(\mu_0
\sigma)^{-1}$. The helical magnetorotational instability (HMRI) was
first discovered theoretically by Hollerbach \& R\"udiger
\cite{HR05}, who realized that adding an azimuthal field $B_{\phi}$
to a vertical field $B_{z}$ can render a differentially rotating
magnetized flow unstable even at very high resistivity up to the
limit $\rm Pm=0$, in contrast to standard magnetorotational
instability (SMRI) with vertical magnetic field \cite{BH}. This
property makes these instabilities amenable to experimental study in
Taylor-Couette (TC) setup filled with low ${\rm Pm}\sim
10^{-6}-10^{-5}$ liquid metals. As a result, the first experimental
detection of HMRI had followed shortly after its theoretical
discovery \cite{Stefani2006}. HMRI was subsequently studied by means
of linear modal stability analysis both with global TC (e.g.,
\cite{Priede2007,Priede2011,Ruediger2008}) and local
(short-wavelength) approaches (see e.g., \cite{LIU,KS2014} and
references therein) as well as via experiments \cite{PRE}. It was
shown that HMRI is determined by the Reynolds number (${\rm Re}$)
and the Hartmann number (${\rm Ha}$), that ensures its persistence
at high resistivity, where SMRI does not normally exist. Using the
local analysis, Liu et al. \cite{LIU} showed that in the presence of
an imposed current-free azimuthal magnetic field HMRI operates for
rotation profiles $\Omega(r)$ with negative or positive shear
steeper than certain critical values. Specifically, this condition,
expressed by the Rossby number ${\rm Ro}= r(2\Omega)^{-1}
d\Omega/dr$, reads as ${\rm Ro}<{\rm Ro}_{\rm LLL}=2(1{-}\sqrt
2)\approx -0.8284$ or ${\rm Ro}>{\rm Ro}_{\rm ULL}=2(1{+}\sqrt
2)\approx 4.8284$, where LLL and ULL refer to the lower and upper
Liu limits, respectively.

The azimuthal magnetorotational instability (AMRI) is a
non-axisymmetric relative of the axisymmetric HMRI that operates for
dominant azimuthal field \cite{TEELUCK} and shares many properties
with the latter. In particular, the same Liu limits define the
threshold of stability for AMRI too \cite{KS2012,KS2014}. The
existence and importance of AMRI for steep positive shear was also
shown recently \cite{SK2015,RUEDIGER}.

Apart from liquid metal flows in laboratory, low-${\rm Pm}$ flows
are found in a wide variety of astrophysical and geophysical
settings: in the ``dead zones'' of protoplanetary disks, in stellar
interiors and in the liquid cores of planets, which thus are the
potential sites for HMRI and AMRI activity. Moreover, in compact
objects, like stars and planets, the condition of decreasing angular
velocity (requirement for SMRI) is not everywhere met, for example,
in the equator-near strip of the solar tachocline \cite{PAME}, which
is also the region of sunspot activity \cite{CHA}.

The Liu limits imply that in the case of current-free field, HMRI
and AMRI do not extend to the astrophysically important Keplerian
rotation with ${\rm Ro}_{Kep.}=-0.75$. To remedy the situation,
Kirillov \& Stefani \cite{KS2013} considered axial electrical
currents not only at the axis, but also in the fluid, resulting in
the azimuthal field $B_{\phi}(r)$ to deviate from the current-free
profile $\propto 1/r$. They generalized the dispersion relation of
Liu et al. \cite{LIU} for this case and derived a new instability
boundary -- a curve in a plane that is spanned by ${\rm Ro}$ and a
corresponding steepness of $B_{\phi}$, called magnetic Rossby
number, ${\rm Rb}=r (2 B_{\phi}/r)^{-1}
\partial{(B_{\phi}/r)}/ \partial r$. In the limit of large ${\rm Re}$ and ${\rm
Ha}$, this curve acquires the closed form
\begin{equation}
{\rm Rb}=-\frac{1}{8}\frac{({\rm Ro}+2)^2}{{\rm Ro}+1}. \label{rel}
\end{equation}
It is seen from this expression that the LLL and the ULL are just
the endpoints of this curve in the current-free regime ${\rm
Rb}=-1$. Condition (1) indicates that even a small axial current
within the liquid can break the lower Liu limit ${\rm Ro}_{\rm LLL}$
and enable HMRI and AMRI to operate for Keplerian profiles. This
effect is now to be investigated in a planned liquid sodium TC
experiment \cite{DRESDYN}, which will combine and enhance the
previous experiments on HMRI \cite{PRE}, AMRI \cite{SEIL2}.

The above-mentioned linear studies of HMRI and AMRI were carried out
in the framework of classical modal stability analysis of fluid
mechanics, which focuses on the behavior at asymptotically large
times. Instead, the nonmodal approach to the stability of shear
flows focuses on the finite-time dynamics of perturbations,
accounting for transient phenomena due to the shear-induced
nonnormality of the flow
\cite{Trefethen92,Schmid_Henningson2001,Schmid2007}. In this
approach, one calculates the optimal initial perturbations that lead
to the maximum possible linear amplification during some finite
time. In self-adjoint flows, the perturbations that grow most are
the least stable solutions of the modal eigenvalue problem. By
contrast, in non-selfadjoint shear flows, the normal mode
eigenfunctions are non-orthogonal due to the nonnormality, resulting
in transient, or nonmodal amplification of perturbations, often by
factors much higher than that of the most unstable normal mode
\cite{Schmid_Henningson2001,Squire2014}. So, leaving out the effects
of the nonnormality can give an incomplete picture of the overall
dynamics (stability) of shear flows.

In this paper, we investigate the nonmodal dynamics of HMRI and AMRI
in differentially rotating magnetized flows, which represent a
special class of shear flows and hence the nonnormality inevitably
plays a role for them. Up to date, these instabilities have been
studied using the modal approach. Recently, the nonmodal dynamics of
SMRI has been addressed by Squire \& Bhattacharjee \cite{Squire2014}
and Mamatsashvili et al. \cite{Mamatsashvili2013}. Here we extend
these investigations to the resistive, or low-${\rm Pm}$ regime,
where only HMRI and AMRI survive. One of our main goals is to link
the magnetohydrodynamic features of these instabilities including
the universal two Liu limits derived with modal approach, which
still remain unexplained, to the nonmodal dynamics of perturbations
in the hydrodynamic case.

\section{Presentation of the problem}

The main equations of non-ideal magnetohydrodynamics for
incompressible conducting media are
\begin{equation}
\frac{\partial {\bf u}}{\partial t}+({\bf u}\cdot \nabla) {\bf
u}=-\frac{1}{\rho}\nabla \left(p+\frac{{\bf B}^2}{2\mu_0}\right)
+\frac{({\bf B}\cdot\nabla) {\bf B}}{\mu_0 \rho} + \nu\nabla^2 {\bf
u},
\end{equation}
\begin{equation}
\frac{\partial {\bf B}}{\partial t}=\nabla\times \left( {\bf
u}\times {\bf B}\right)+\eta\nabla^2{\bf B},
\end{equation}
\begin{equation}
\nabla\cdot {\bf u}=0,~~~\nabla\cdot {\bf B}=0,
\end{equation}
where $\rho=const$ is the density, $\nu=const$ the kinematic
viscosity, and $\eta$ the magnetic diffusivity, $p$ is the thermal
pressure, ${\bf u}$ is the velocity and ${\bf B}$ is the magnetic
field.

An equilibrium flow represents a fluid rotating with angular
velocity $\Omega(r)$ and threaded by a magnetic field, which
comprises a constant axial component $B_{0z}$ and an azimuthal one
$B_{0\phi}$ with arbitrary radial dependence:
\[
{\bf u}_0=r\Omega(r){\bf e}_{\phi}, ~~~~{\bf B}_0=B_{0\phi}(r){\bf
e}_{\phi}+B_{0z}{\bf e}_z.
\]

Consider now small perturbations about this equilibrium, ${\bf
u}'={\bf u}-{\bf u}_0$, $p'=p-p_0$, ${\bf B}'={\bf B}-{\bf B}_0$.
Following \cite{KS2014} we adopt a local (Wentzel-Kramers-Brillouin,
WKB) approximation in the radial coordinate around some fiducial
radius $r_0$, i.e., assume perturbation length-scales to be much
shorter than the characteristic length of radial variations of the
equilibrium quantities, and represent the perturbations as ${\bf
u}',{\bf B}'\propto \exp({\rm i}k_rr+{\rm i}m\phi+{\rm i}k_zz)$,
with azimuthal $m$, axial $k_z$ and large radial $k_r$ wavenumbers,
$r_0k_r\gg1$ (without loss of generality we take $m,k_z>0$).
Linearizing Eqs. (2)-(4) about the equilibrium, introducing new
variables $\xi={\rm i}(k_zu'_{\phi}-mu'_z/r_0)$ and
$\zeta=i(k_zB'_{\phi}-mB'_z/r_0)$ as in \cite{Squire2014}, and
normalizing time by $\Omega^{-1}$, distance by $r_0$ and velocity by
$r_0\Omega$, we arrive at the following equations for the
perturbations in nondimensional form (primes are omitted and the
factor $(\mu_0\rho)^{-1/2}$ is absorbed in the magnetic field)
\cite{KS2014,Squire2014}
\begin{equation}
\frac{d\boldsymbol \psi}{dt}={\bf A}\cdot {\boldsymbol \psi},
\end{equation}
where ${\boldsymbol \psi}\equiv (u_r, \xi, B_r, \zeta$) is the state
vector and the evolution matrix operator ${\bf A}$ is
\[
{\bf A}=\begin{pmatrix}
-\frac{k^2}{\rm Re}+4{\rm Ro}\frac{mk_r}{k^2} & -2{\rm i}\frac{\alpha}{k} & {\rm i}F &  2{\rm i}\omega_{\phi}\frac{\alpha}{k} \\
-2{\rm i}(1+{\rm Ro})k_z & -\frac{k^2}{\rm Re} & 2{\rm
i}\omega_{\phi}(1+{\rm Rb})k_z & {\rm
i}F \\
 {\rm i}F & 0 & -\frac{k^2}{\rm Rm} & 0 & \\
-2{\rm i}\omega_{\phi}{\rm Rb}\cdot k_z & {\rm i}F & 2{\rm i}{\rm
Ro}\cdot k_z & -\frac{k^2}{\rm Rm},
\end{pmatrix}
\]
where $k_r(t)=k_{r}(0)-2{\rm Ro}\cdot mt$, $k^2=k_r^2+m^2+k_z^2$,
$\alpha=k_z/k$ and $F= m\omega_{\phi}+\omega_z$ with $\omega_{\phi}=
B_{0\phi}/r_0\Omega,~\omega_z=k_zB_{0z}/\Omega$. Note that for
non-axisymmetric ($m \neq 0$) perturbations the radial wavenumber
$k_r$ varies with time due to the advection by the background flow.
The Reynolds number, ${\rm Re}=\Omega r_0^2/\nu$, and the magnetic
Reynolds number, ${\rm Rm}=\Omega r_0^2/\eta$, are fixed to ${\rm
Re}=4000$ and ${\rm Rm}=0.0056$, yielding a small magnetic Prandtl
number ${\rm Pm}={\rm Rm}/{\rm Re}=1.4\cdot 10^{-6}$ typical for
liquid metals. The strength of the imposed field is measured by the
Hartmann number ${\rm Ha}=B_0r_0/\sqrt{\nu \eta}$, which is in the
range ${\rm Ha}\sim 10-100$ in liquid metal experiments
\cite{PRE,SEIL2}. The relative effect of the azimuthal magnetic
field over the axial one is characterized by the ratio
$\beta=\omega_{\phi}/\omega_z$. The HMRI and AMRI are driven by the
terms proportional to $\omega_{\phi}$ in Eq. (5) and hence are
effective in the presence of an appreciable azimuthal field,
respectively, for $\beta \sim 1$ and $\beta \gg 1$ (see e.g.,
\cite{LIU,KS2014}). We consider Rayleigh-stable rotation with ${\rm
Ro}> -1$, and ${\rm Rb}<0$, since the axial current decreases with
radius. It is readily shown that ${\bf A}$ is indeed nonnormal,
i.e., it does not commute with its adjoint, ${\bf A}^{\dag}\cdot
{\bf A}\neq {\bf A}\cdot {\bf A}^{\dag}$.

We quantify the nonmodal amplification in terms of the total
perturbation energy, $E=\frac{\rho}{2}(|{\bf u}|^2+|{\bf
B}|^2)={\boldsymbol \psi}^{\dag}\cdot F^{\dag}F\cdot{\boldsymbol
\psi}$, where ${\bf F}=\sqrt{\rho/2}\cdot
diag(\alpha^{-1},1,\alpha^{-1},1)$, which is a physically relevant
norm. The maximum possible, or optimal growth at a specific time $t$
is defined as the ratio $G(t)=\max_{\boldsymbol \psi(0)} E(t)/E(0)$,
where $E(t)$ is the energy at $t$ and the maximization is done over
all initial states ${\boldsymbol \psi}(0)$ with a given initial
energy $E(0)$ (e.g., Ref. \cite{Schmid_Henningson2001}). The final
state at $t$ is found from the initial state at $t=0$ by solving the
linear Eq. (5) and can be formally written as ${\boldsymbol
\psi}(t)={\bf K}(t)\cdot {\boldsymbol \psi}(0)$, where ${\bf K}(t)$
is the propagator matrix. Then, the maximum possible amplification
$G(t)$ is usually calculated by the singular value decomposition of
${\bf K}$ at $t$ (e.g., Refs. \cite{Schmid_Henningson2001}). The
square of the largest singular value gives the value of $G(t)$ and
the corresponding initial condition that leads to this growth,
optimal perturbation, is given by the right singular vector. We
stress again that the nonmodal approach combined with the method of
optimal perturbations is the most general way of analyzing shear
flow dynamics (stability) at all times, as opposed to the modal
approach, which is concerned only with the behavior at asymptotic
times and hence omits important finite-time phenomena.


\section{Dispersion relation}

Before embarking on investigating the nonmodal dynamics of HMRI and
AMRI, we briefly recap the results from the modal analysis of these
instabilities in the local approach \cite{LIU,Priede2011,KS2014}. In
this case, $|k_r(t)|\gg m$ or $k_z\gg m$ and as a result the
shear-related term proportional to $m$ in $A_{11}$, $4{\rm Ro}\cdot
mk_r/k^2$, as well as the time-dependence of the radial and total
wavenumbers are ignored, $k_r'(t)/|{\rm Ro}|\ll k_r(t)$,
$k'(t)/|{\rm Ro}|\ll k(t)$. This admits WKB approach in \emph{time}
when solution can be sought in the form $\propto \exp(-{\rm i}\int
\omega(t')dt')$, with the adiabatic condition $\omega'(t)\ll
\omega^2(t)$ being fulfilled. Substituting this into Eq. (5) and
taking the relevant limit of small magnetic Reynolds number, ${\rm
Rm}\ll 1$ (inductionless approximation), but high Reynolds number,
${\rm Re}\rightarrow \infty$, we arrive at the following analytical
expression for the growth rate $\gamma=Im(\omega)$ \cite{KS2014},
\begin{equation}
\gamma=(2\alpha^2\omega_{\phi}^2\cdot {\rm Rb}-F^2)\frac{\rm
Rm}{k^2}-\frac{k^2}{\rm Re}+\sqrt{2X+2\sqrt{X^2+Y^2}},
\end{equation}
where
\[
X=\alpha^2\omega_{\phi}^2(\alpha^2\omega_{\phi}^2\cdot {\rm
Rb}^2+F^2)\frac{{\rm Rm}^2}{k^4}-({\rm Ro}+1)\alpha^2,
~~~Y=\omega_{\phi}\alpha^2F({\rm Ro}+2)\frac{\rm Rm}{k^2}.
\]
An instability occurs when the growth rate is positive, $\gamma>0$.

Now consider the cases of HMRI and AMRI. HMRI relies on the growth
of axisymmetric ($m=0$) perturbations and appears from ${\rm Ha}\sim
10$. Taking the limit of small interaction parameter, ${\rm
Ha^2/Re}\ll 1$ and then maximizing with respect to $\beta$ [$\beta
\sim O(1)$], Eq. (6) simplifies to
\begin{equation}
\gamma=\alpha^2\frac{{\rm Ha}^2}{\rm Re}\left[\frac{({\rm
Ro}+2)^2}{8({\rm Ro}+1)(-{\rm Rb})}-1\right],
\end{equation}
(here ${\rm Ha}$ is defined in terms of $B_{0z}$, ${\rm
Ha}=B_{0z}r_0/\sqrt{\nu \eta}$). On the other hand, AMRI consists in
the growth of non-axisymmetric perturbations and takes place when
the azimuthal magnetic field dominates over axial one, corresponding
to the limit $\beta \rightarrow \infty$ in Eq. (6)
\begin{equation}
\gamma=(2\alpha^2\cdot{\rm Rb}-m^2)\frac{1}{k^2}\frac{\rm Ha^2}{\rm
Re}-\frac{k^2}{\rm Re}+\sqrt{2X+2\sqrt{X^2+Y^2}},
\end{equation}
\[
X=(\alpha^2\cdot {\rm Rb}^2+m^2)\frac{\alpha^2}{k^4}\frac{{\rm
Ha}^4}{\rm Re^2}-({\rm Ro}+1)\alpha^2, ~~~Y=m({\rm
Ro}+2)\frac{\alpha^2}{k^2}\frac{\rm Ha^2}{\rm Re},
\]
where now ${\rm Ha}$ is appropriately defined in terms of
$B_{0\phi}$, ${\rm Ha}=B_{0\phi}r_0/\sqrt{\nu \eta}$. If the
interaction parameter is small, ${\rm Ha^2/Re}\ll 1$, and ${\rm
Re}\rightarrow \infty$, Eq. (8) reduces to
\[
\gamma=\frac{1}{k^2}\frac{\rm Ha^2}{\rm Re}\left(2\alpha^2\cdot {\rm
Rb}-m^2+m\alpha\frac{{\rm Ro}+2}{\sqrt{{\rm Ro}+1}}\right).
\]
To the leading order in ${\rm Rm}$, the corresponding real part of
the eigenfrequency is equal to the frequency of inertial waves (with
minus sign), $Re(\omega)=-\omega_{\rm iw}=- 2\alpha\sqrt{1+{\rm
Ro}}$. Remarkably, both Eqs. (7) and (8) yield the same stability
boundary (1) defined by $\gamma=0$ \cite{KS2014}, which indicates
that for the current-free field, ${\rm Rb}=-1$, modal growth of HMRI
and AMRI exists at negative shear less than the lower Liu limit,
${\rm Ro} < {\rm Ro}_{\rm LLL}=-0.8284$ and at positive shear larger
than the upper Liu limit, ${\rm Ro}
> {\rm Ro}_{\rm ULL}=4.8284$, while at larger ${\rm Rb}>-1$ the
stability region shrinks and the instability extends inside the Liu
limits. So, the modal growth of HMRI and AMRI can, in principle,
also exist for the Keplerian rotation (${\rm Ro}_{\rm Kep.}=-0.75$)
starting from ${\rm Rb}=-0.781$ \cite{KS2013}. From Eq. (8) it
follows that in these intervals of Rossby numbers, AMRI operates
(i.e., $\gamma>0$) only in a certain range of radial wavenumbers,
outside this range $\gamma<0$ and the perturbations decay due to
resistivity. Since we are interested in the effects of nonnormality
on the dynamics of HMRI and AMRI, below we focus on the current-free
azimuthal field, i.e., fix ${\rm Rb}=-1$, where free energy for
instability comes solely from shear.

\section{Nonmodal dynamics of HMRI}

Now, following Ref. \cite{MS2016}, we investigate the nonmodal
growth of axisymmetric HMRI by solving an initial value problem
given by Eq. (5) as described in Sec. 1, without restricting the
time-dependence of harmonics (modes) to the exponential form, as
accepted in modal analysis. In the case of axisymmetric HMRI, $k_r$
does not vary with time and hence the evolution matrix ${\bf A}$ is
stationary. Figure 1 shows the maximum energy growth $G(t)$ at
modally stable and unstable ${\rm Ro}$ together with the growth in
the modally stable nonmagnetic case, where only the nonmodal growth
is possible. For HMRI we take ${\rm Ha}=15$. In all cases, the
initial stage of evolution is qualitatively similar: the energy
increases with time, reaches a maximum $G_m$ and then decreases.
This first nonmodal amplification phase is followed by minor
amplifications. Like in the case of modal growth, the kinetic energy
dominates over the magnetic one also during nonmodal growth. As a
result, the duration of each amplification event is set by inertial
waves: the peak value $G_m$ is attained at around one quarter of the
wave period, $t_m\approx \pi/2\omega_{\rm iw}$, similar to that in
the nonmagnetic case, although its value is smaller than that in the
latter case. At larger times, the optimal growth follows the
behavior of the modal solution -- it increases (for ${\rm Ro}=-0.9,
7$), stays constant (for the Liu limits, ${\rm Ro}={\rm Ro}_{\rm
LLL}, {\rm Ro}_{\rm ULL}$) or decays (for ${\rm Ro}=-0.75,-0.6, 2$),
respectively, if the flow is modally unstable, neutral or stable; in
the latter case HMRI undergoes only transient amplification. This is
readily understood: at large times the least stable modal solution
(with growth rate given by Eq. 7) dominates, whereas at small and
intermediate times the transient growth due the interference of
nonorthogonal eigenfunctions is important. In particular, for the
Liu limits, where the modal growth is absent, there is still
moderate nonmodal growth $G_m({\rm Ro_{LLL}})=4.06, G_m({\rm
Ro_{ULL}})=5.46$. 
A similar evolution of axisymmetric perturbations' energy with time
for HMRI was already found in \cite{Priede2007}, where also the
physical mechanism of HMRI was explained in terms of an additional
coupling between meridional and azimuthal flow perturbations.
Importantly, in Fig. 1, $G_m$ at modally stable and unstable Rossby
numbers are comparable and several times larger than the modal
growth factors during the same time $t_m$. Indeed, for example, at
${\rm Ro}=-0.9$ the growth achieves the first peak $G_m=5.71$ at
$t_m=2.2$, while at this time the energy of the normal mode would
have grown only by a factor of $\exp[2t_m\gamma({\rm Ro})]=1.135$.
This also implies that in the Keplerian regime, where there is no
modal growth of HMRI for ${\rm Rb} = -1$, it still exhibits moderate
nonmodal growth (red curve in Fig. 1a).

Figure 2, which is the central result of this paper, shows the
maximum growth $G_m$ in the magnetic and nonmagnetic cases together
with the modal growth rate $\gamma$ given by Eq. (7) versus ${\rm
Ro}$. $G_m$ increases linearly with ${\rm Ro}$ at ${\rm Ro}>0$ and
much steeper at ${\rm Ro}<0$ which can be well approximated by
$\propto(1+{\rm Ro})^{-0.78}$. For comparison, in this plot we also
show the maximum transient growth factor for axisymmetric
perturbations in the nonmagnetic case, $G_m^{(h)}=(1+{\rm Ro})^{{\rm
sgn}({\rm Ro})}$, from \cite{Afshordi2005}. So, although $G_m$ in
the magnetic case is slightly smaller than that in the nonmagnetic
one, the two curves are in fact close to each other and display
nearly the same behavior with ${\rm Ro}$. Note that the dependencies
of $G_m$, $G_m^{(h)}$ (Fig. 2a) and of the modal growth rate
$\gamma$ (Fig. 2b) on ${\rm Ro}$ have very similar shapes.
Remarkably, the latter, being given by Eq. (7), can be expressed in
terms of the hydrodynamic nonmodal growth $G_m^{(h)}=(1+{\rm
Ro})^{{\rm sgn}({\rm Ro})}$ in the closed form
\begin{equation}
\gamma=\frac{{\rm Ha}^2}{\rm Re}\left[\frac{(G_m^{(h)}+1)^2} {8
G_m^{(h)}}-1 \right] \label{connection}
\end{equation}
which is indeed proportional to $G_m^{(h)}$ for larger values. Both
Liu limits are therefore connected with a corresponding threshold
$G_m^{(h)}({\rm Ro}_{\rm LLL})=G_m^{(h)}({\rm Ro}_{\rm ULL})=5.828$.

\section{Nonmodal dynamics of AMRI}

For non-axisymmetric AMRI, the radial wavenumber changes with time
due to background shear and as a result its dynamics differs from
that of axisymmetric HMRI. The adiabatic WKB regime in time applies
only at $|k_r(t)|\gg m$ ($k_z\lesssim m$), then $|k_r|$ decreases
with time and enters the non-adiabatic interval in the neighborhood
of $|k_r(t)| \sim m$ where the dispersion relation (8) and hence the
modal approach are no longer applicable for the description of AMRI.
In this case, one should resort to numerical integration of Eq. (5)
to study the dynamics of modes and quantify their nonmodal
amplification. It is in this non-adiabatic region where the effects
of nonnormality manifest themselves and influence the dynamics of
AMRI. Obviously, the modes with those initial $k_r(0)$ which do not
lead to crossing the non-adiabatic interval, i.e., with $k_r(0)$ and
${\rm Ro}\cdot m$ having opposite signs, do not experience the
transient growth and decay quickly due to resistivity. Here, we
restrict ourselves to most amplified $m=1$ modes, while the dynamics
of other modes with larger $m$, which usually grow less than the
$m=1$ modes do, will be presented elsewhere. Figure 3 shows the
evolution of the maximum energy growth $G(t)$ for these harmonics at
various ${\rm Ro}$, including the Keplerian rotation, and fixed
$k_z=m=1$. The function $G$ has been maximized over the initial
wavenumber $k_r(0)$, which is negative (positive) at ${\rm Ro}<0$
(${\rm Ro}>0$) with the absolute value $|k_r(0)|\gg m$. The Hartmann
number characterizing the azimuthal field is set to ${\rm Ha}=100$,
which is typical for AMRI \cite{SEIL2}. Initially, in the adiabatic
region, the effect of resistivity on the mode is still appreciable.
As the mode evolves, $|k_r(t)|$ decreases, resistive dissipation
becomes weaker, while the effect of nonnormlity/shear gets stronger.
As a result, the energy starts to amplify, extracting energy from
the background flow. Then, $|k_r(t)|$ decreasing further, enters the
non-adiabatic region, where the effect of the nonnormality is
largest. As a consequence, $G$ exhibits most of growth just in this
interval of radial wavenumbers, reaching a maximum, $G_m$, at
different $|k_{r,m}|\lesssim 2$, which depend on ${\rm Ro}$, but are
close to each other. The peak $G_m$ is higher, the larger is the
shear $|{\rm Ro}|$ (see also Fig. 4). Afterwards, $|k_r|$ increases
again, leaving the non-adiabatic area, and the harmonic's energy
gradually decreases and eventually decays due to viscosity and
resistivity at high enough $k_r$. We refer to this process as the
nonmodal growth of AMRI, which always lasts for a finite time in the
local approach, because of the shear-induced time variation of the
radial wavenumber of non-axisymmetric modes. Note that, like in the
case of HMRI, moderate nonmodal growth occurs also at the Liu limits
and the Keplerian rotation, which are, respectively, marginally
stable and fully stable according to modal analysis.

As mentioned above, the optimal perturbation and optimal nonmodal
growth formalism, as opposed to the modal approach, is the most
general way to describe the dynamics of non-axisymmetric
perturbations. In the adiabatic regime, for HMRI and AMRI it gives
the results of the modal analysis, but for the latter one should
also take into account the time-dependence of the radial wavenumber
and neglect small in this regime shear-induced terms in Eq. (5)
(second term in $A_{11}$). The situation in this highly resistive
flow is analogous to the nonmodal (transient) growth of
non-axisymmetric perturbations in modally/spectrally stable
unmagnetized shear flows \cite{Mukho2005,Maretzke2014}, except that
with the imposed azimuthal magnetic field there exists an additional
means of energy gain from the mean flow due to the terms
proportional to $\omega_{\phi}$ in main Eq. (5), i.e., AMRI, whose
dynamics itself is modified by the nonnormality. By contrast, for
axisymmetric perturbations (i.e., for HMRI), as seen above, the
nonmodal amplification precedes the modal growth.

Figure 4 shows the dependence of $G_{m}$ on the vertical wavenumber
$k_z$ at several negative and positive Rossby numbers. At larger
absolute values of ${\rm Ro}$, ${\rm Ro}=-0.92, -0.9$ and 7, it
first increases at small $k_z$, achieves a peak at $k_{z,m}\sim 1-2$
and then decreases at large $k_z$, more rapidly for positive shear.
The critical wavenumber $k_{z,m}$ decreases with $|{\rm Ro}|$, and
eventually becomes $k_{z,m}=0$. It is seen in Fig. 4 that the
nonmodal growth is more than an order of magnitude larger at
positive ${\rm Ro}$ than at negative ${\rm Ro}$, indicating the
importance of positive shear for AMRI, as was already shown recently
in \cite{SK2015}, but using the modal approach. Note that
$k_{z,m}\sim m$ and hence falls in the non-adiabatic regime,
implying that the nonmodal approach is more appropriate to describe
the dynamics of AMRI rather than modal one at the these vertical
wavenumbers that yield the maximum growth. At $k_z\gg m$, the
temporal WKB approximation holds at all times during evolution
(except near points $\gamma(k_r)=0$), if any). Due to our general
treatment of an initial value problem posed by Eq. (5), the growth
factor at large $k_z$ in Fig. 4 essentially coincides with that
given by modal analysis.

\section{Conclusions}

In this paper, we investigated the linear nonmodal dynamics of HMRI
and AMRI due to the nonnormality of shear magnetized flow with large
resistivity in the local approximation. As a main tool of analysis
we used nonmodal approach in combination with the optimal
perturbation formalism, which allow to characterize the growth of
perturbations in the most general form, comprising also the modal
regime. We traced the entire time evolution of modes by solving an
initial value problem, thereby capturing the finite-time dynamics.
As shown in Fig. 2 and quantified exactly in Eq. (9), the modal
growth rate of HMRI exhibits a very similar dependence on ${\rm Ro}$
as the maximum nonmodal growth in the purely hydrodynamic shear
flow, establishing a fundamental link between nonmodal dynamics and
dissipation-induced modal instabilities, such as HMRI. Both, despite
the latter being of magnetic origin, rely on hydrodynamic means of
amplification, i.e., extract energy from the background flow mainly
by Reynolds stress due to shear/nonormality \cite{Priede2007}. The
dynamics of AMRI is more complex due to the shear-induced
time-variation of the radial wavenumber of non-axisymmetric modes.
As a result, in the local approach, the growth of AMRI both in modal
(adiabatic) and nonmodal (non-adiabatic) regimes is always
transient. The maximum nonmodal growth factor increases with shear,
i.e., with the absolute value of Rossby number and achieves an order
of magnitude higher values at positive shear than at negative shear,
consistent with the recent findings of \cite{SK2015} on the
existence and importance of AMRI for positive shear.

\section*{Acknowledgements}

This work was supported by the Alexander von Humboldt Foundation and
the Helmholtz Association in frame of the Helmholtz Alliance
LIMTECH. Discussions with M. Avila and A. Guseva are gratefully
acknowledged.



\newcommand{\noopsort}[1]{}

\clearpage

\begin{figure}
\centering
\includegraphics[width=\textwidth, height=0.6\textwidth]{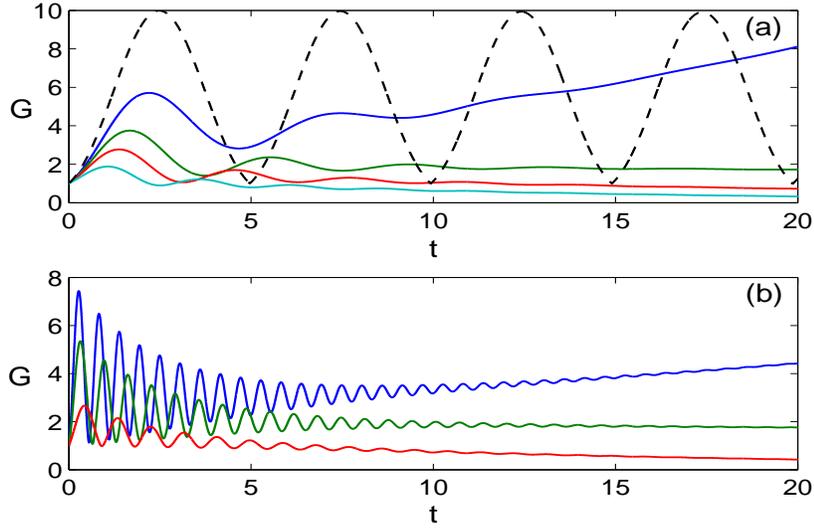}
\caption{$G(t)$ for HMRI with $m=0$ at different (a) ${\rm Ro}=-0.9
(blue)$, $-0.8284 ({\rm LLL}, green)$, $-0.75 ({\rm Kepler}, red)$,
$-0.6 (cyan)$ and (b) ${\rm Ro}=2 (red)$, $4.8284 ({\rm ULL},
green)$, $7 (blue)$. For reference, the dashed black curve in panel
(a) shows the maximum growth factor vs. time in the nonmagnetic case
at ${\rm Ro}=-0.9$. The other parameters are $\alpha=1$ and $k=1$.
For each ${\rm Ro}$, the parameter $\beta$ is chosen such that to
maximize the modal growth rate for given other parameters.}
\end{figure}

\begin{figure}
\centering
\includegraphics[width=\textwidth, height=0.6\textwidth]{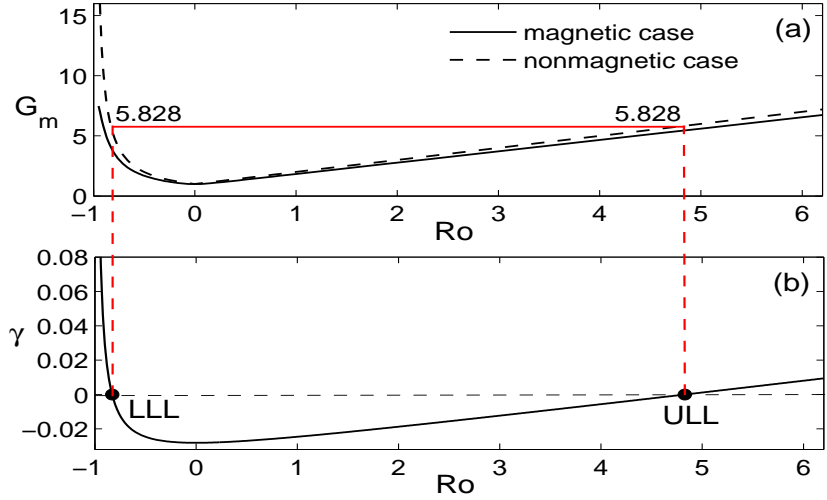}
\caption{Shown are (a) $G_m$ for HMRI (solid line) and for the
nonmagnetic case with (dashed line) as well as (b) the modal growth
rate of HMRI from Eq. (7) versus ${\rm Ro}$. Other parameters are as
in Fig. 1. Red lines illustrate connection between the Liu limits of
HMRI and the nonmodal growth in the purely hydrodynamic case.}
\end{figure}

\begin{figure}
\includegraphics[width=\textwidth,height=0.6\textwidth]{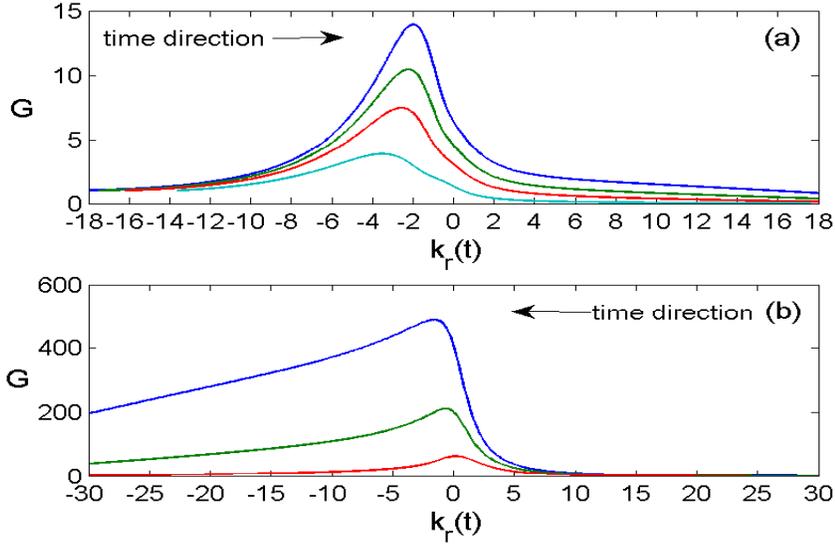}
\caption{$G(t)$ as a function of time-dependent $k_r(t)$ for AMRI
with $m=1$ and $k_z=1$ at different negative Rossby numbers (a)
${\rm Ro}=-0.9 (blue), -0.8283 ({\rm LLL},green), -0.75
(Kepler,red), -0.6 (cyan)$, where $k_r$ increases with time from
negative to positive values, and at positive Rossby numbers (b)
${\rm Ro}=3 (red), 4.8284 ({\rm ULL},green), 7 (blue)$, where $k_r$
decreases with time from positive to negative values. Each curve has
been maximized with respect to the initial value of the radial
wavenumber $k_r(0)$. The larger is shear $|{\rm Ro}|$, the higher is
the growth.}
\end{figure}

\begin{figure}
\includegraphics[width=\textwidth,height=0.5\textwidth]{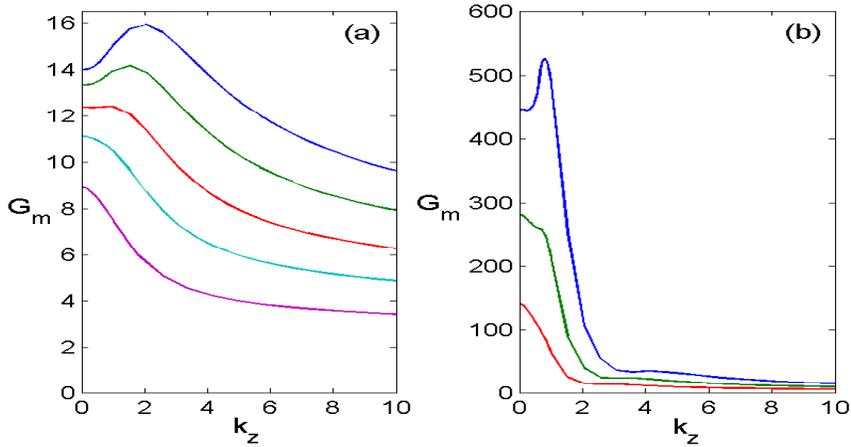}
\caption{$G_m$ vs. $k_z$ for $m=1$ modes at various Rossby numbers
(a) ${\rm Ro}=-0.92 (blue), -0.9 (green), -0.87 (red), -0.8284 ({\rm
LLL},cyan) -0.75 (Kepler, violet)$ and (b) ${\rm Ro}=3 (red), 4.8284
(${\rm ULL}$, green), 7 (blue)$. For positive shear (b), the
nonmodal growth is more than an order of magnitude larger than that
for negative one.}
\end{figure}

\lastpageno
\end{document}